\documentclass[9pt,twocolumn,twoside]{optica}
\setboolean{shortarticle}{false}
\setboolean{minireview}{false}

\title{Twin-beam intensity-difference squeezing below 10~Hz}

\author[1,2]{Meng-Chang Wu}
\author[1,3]{Bonnie L. Schmittberger}
\author[1,4]{Nicholas R. Brewer}
\author[1]{Rory W. Speirs}
\author[5]{Kevin M. Jones}
\author[1,2,4*]{Paul D. Lett}

\affil[1]{Joint Quantum Institute, National Institute of Standards and Technology and the University of Maryland, College Park, Maryland 20742, USA}
\affil[2]{Chemical Physics Program, University of Maryland, College Park, Maryland 20742, USA}
\affil[3]{Present Address: Physical Sciences, Nanosystems and Quantum Group, The MITRE Corporation, Princeton, NJ 08540, USA}
\affil[4]{Quantum Measurement Division, National Institute of Standards and Technology, Gaithersburg, Maryland 20899, USA}
\affil[5]{Department of Physics, Williams College, Williamstown, Massachusetts 01267, USA}

\affil[*]{Corresponding author: paul.lett@nist.gov}

\dates{Compiled \today}

\ociscodes{(270.6570) Squeezed states; (270.0270) Quantum optics; (190.4380) Nonlinear optics, four-wave mixing.}

\doi{\url{ }}

\begin{abstract}
We report the generation of strong, bright-beam intensity-difference squeezing down to measurement frequencies below 10 Hz. We generate two-mode squeezing in a four-wave mixing (4WM) process in Rb vapor, where the single-pass-gain nonlinear process does not require cavity locking and only relies on passive stability. We use diode laser technology and several techniques, including dual seeding, to remove the noise introduced by seeding the 4WM process as well as the background noise. Twin-beam intensity-difference squeezing down to frequencies limited only by the mechanical and atmospheric stability of the lab is achieved. These results should enable important low-frequency applications such as direct intensity-difference imaging with bright beams on integrating detectors.
\end{abstract}

\setboolean{displaycopyright}{false}

\begin{document}

\maketitle
\thispagestyle{fancy}
\ifthenelse{\boolean{shortarticle}}{\abscontent}{}

\section{Introduction}

Quantum-enhanced sensing technologies have become increasingly important in a number of fields as the limitations of classical technologies are approached. It will be important for certain applications to  perform sub-shot-noise measurements at low frequencies. In particular, gravity-wave interferometers~\cite{ref1, ref2} operate in the audio frequency range of 10 Hz to 10 kHz and presently employ single-mode quadrature squeezing, but a number of other potential low-frequency applications of squeezing are possible. These include quantum information storage in optical memories based on electromagnetically-induced transparency, opto-acoustic or thermo-optical spectroscopic techniques~\cite{ref3, ref4}, as well as magnetometry~\cite{ref5}. Detector calibration techniques involving squeezed light~\cite{ref6, ref7} would also benefit from operation at frequencies as low as 100 Hz, as this is where current metrological calibrations are performed. Direct intensity-difference measurements on a CCD camera are also an important goal. Being able to make such measurements with relatively portable and affordable sources, such as those based on diode laser systems, is  important for applications outside of a laboratory setting. 

In this work we show how to extend the generation of bright, two-mode, intensity-difference squeezed light from 4-wave mixing (4WM) in Rb vapor to low frequencies using a diode-laser-based system. In particular, we introduce a dual-seeding technique where two complementary 4WM processes are seeded in order to balance the excess noise due to the seed beams themselves. The present results imply that the squeezing produced by this 4WM process has been previously limited at low frequencies only by technical noise, and thus single-mode quadrature squeezing based on the same 4WM interaction ought to be extendable down to a similar frequency range.

Twin-beam generation, which produces two entangled beams of light, has long held the promise of being able to bring sub-shot-noise measurements to a variety of applications. A technique called quantum-dense metrology involves measuring two orthogonal quadratures of a frequency-degenerate two-mode squeezed state in order to discriminate against noise leakage between quadratures and remove common-mode classical noise~\cite{ref8, ref9}, and has been proposed for use in gravitational-wave detectors. A direct application to interferometry has also been proposed~\cite{ref10}. Differential-absorption imaging~\cite{ref11} allows sub-shot noise signals to be extracted from the difference of two highly-correlated images. A particularly simple twin-beam generation technique, 4-wave mixing in alkali vapors, has always seemed promising in this regard because the squeezing can be generated in many spatial modes. Observing such squeezing on an integrating detector, like a CCD camera, requires that the light be squeezed when integrated over the entire frequency range of the measured pulse of light, and holds the promise of demonstrating a real quantum sensitivity advantage for a camera array with fixed well-depth pixels. Unfortunately, technical noise at low frequencies has proven to be a severe limitation in this regard.  

Generating squeezed light at low ($<20$ kHz) or acoustic measurement frequencies has been pursued in a number of different systems and forms. The lowest frequency squeezing has been observed in single-mode quadrature squeezing, but 2-mode squeezing and polarization squeezing have also been pushed down into the acoustic frequency range. Single-mode quadrature squeezing has perhaps its most important applications in future gravity-wave interferometers~\cite{ref12, ref13, ref14}, and such squeezing at frequencies as low as 1 Hz has been observed from an optical parametric oscillator (OPO) with this application in mind~\cite{ref14, ref15}. Squeezing has not been reported at such low frequencies for two-mode intensity-difference measurements, and the best results seem to be $\approx1.5$~kHz reported in~\cite{ref16}, and $\approx700$~Hz recently reported in~\cite{ref17}. 

Obtaining squeezing at low frequencies is generally not an easy task, and in the single-mode-squeezing case the above-mentioned results were obtained using a complex system employing two additional frequency-shifted control beams to sense and to feed back on the OPO cavity length and the pump phase. In contrast, the results presented here demonstrate intensity-difference squeezing down to $\approx10$~Hz using an extremely simple, passive optical system (in the sense that there is no active feedback to the optical system beyond the stabilization of the seed laser).

Early OPO systems were constructed with $\chi^{(2)}$ nonlinear crystals in external cavities that enhance the pump field as well as the down-converted fields. These cavities resulted in a high sensitivity to acoustic noise, which limited the squeezing at these frequencies. The use of semimonolithic and monolithic cavity designs, incorporating the cavity into the nonlinear crystal, substantially reduced this sensitivity to noise, and with feedback locking eventually led to the 1 Hz results discussed above. Broadband spontaneous parametric downconversion (PDC) systems that are pumped without cavities have been reported where twin-beam differencing can be performed on an integrating detector, but the photon flux per mode is usually low, requiring a relatively high-power pulsed laser system. For imaging purposes PDC has potentially very high resolution, with many spatial modes. Correlations in the speckle patterns, pixel-by-pixel correlations, and imaging have been demonstrated with these systems~\cite{ref18, ref19, ref20, ref21}. Recently Samantaray et al.~\cite{ref22} worked with continuous-wave pumping and 100 ms integration times to demonstrate an improved imaging system. An important feature that is common to each of these experiments is that there is no seed beam present, which eliminates the noise associated with the seed.

A different type of squeezing, polarization squeezing, or a squeezing of the Stokes parameters, has also been experimentally observed at low frequencies. This is, in particular, useful for magnetometry applications and has been pushed down to the kHz regime~\cite{ref23}, and to $\approx$ 200 Hz in~\cite{ref24}. This particular type of squeezing, however, does not lend itself to the imaging applications that we are interested in.

Previous experiments have reported twin-beam intensity-difference squeezing from 4WM down into the kHz range~\cite{ref16, ref25, ref26}. Very recently Ma et al.~\cite{ref17} reported such squeezing down to frequencies below 700 Hz. In the context of imaging, Kumar et al.~\cite{ref27} performed intensity-differencing on an EMCCD camera using a 4WM source similar to ours and were able to obtain a subtraction noise level 2 dB below shot noise. In those experiments two frames are taken in rapid succession for each of the twin images and are subtracted. This results in images that contain the intensity fluctuations of each beam and removes much of the low-frequency classical noise. 

Producing bright twin beams typically requires seeding the beams, but the seeding itself often limits the low-frequency squeezing performance. Here we present a number of improvements to our established 4WM optical set-up that allow bright, low-frequency intensity-difference squeezing to be observed down to frequencies that are limited by the laboratory beam-pointing stability. We show that we can balance the noise on the seed beam itself by creating two ``complementary'' seeded 4WM processes that balance the seeds on the two detectors. In addition, in this context we demonstrate the benefits of frequency-narrowing the source laser, and of increasing the phase matching angle. While a number of the techniques introduced here are also applicable to systems based on intrinsically less-noisy technologies, such as Ti:sapphire lasers, we use semiconductor diode laser and tapered amplifier technologies. It is important for practical uses that these more portable and affordable technologies work at these frequencies as well.

\begin{figure*}
\centering
\includegraphics[width=0.9\linewidth]{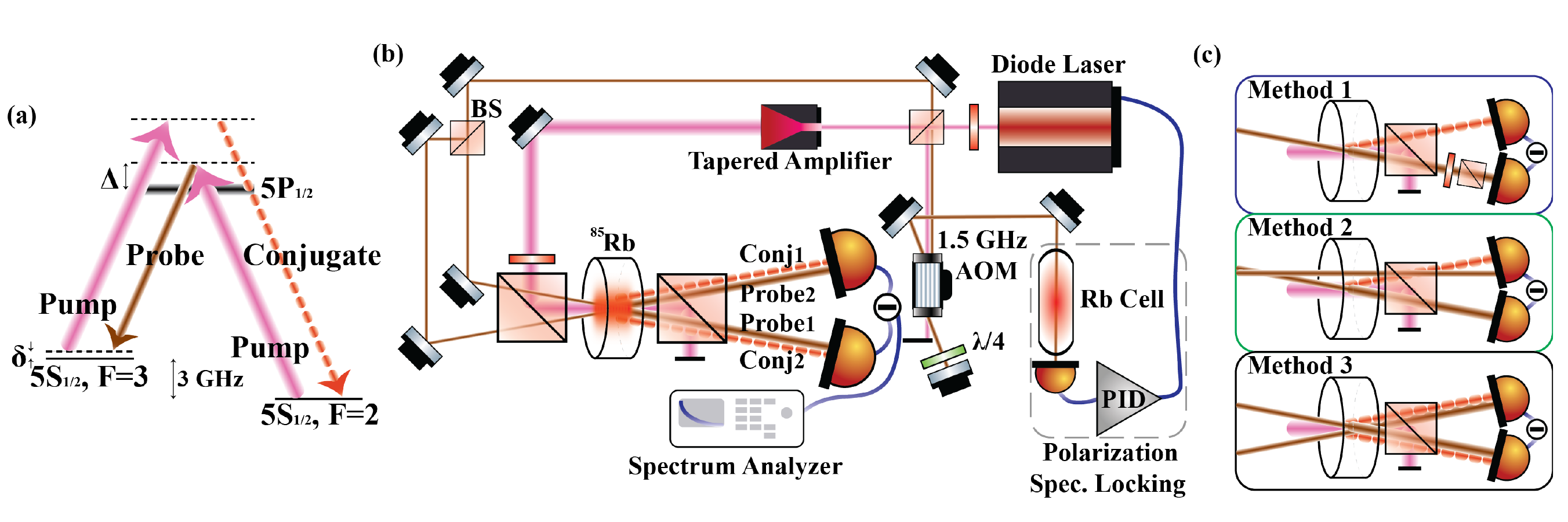}
\caption{a) Level diagram of the 4WM process in $^{85}$Rb. The broadened upper level indicates the Doppler broadening of the transition. b) Experimental set-up. BS indicates a non-polarizing beamsplitter; the remaining beamsplitters are polarizing beamsplitters.  The dashed box indicates a polarization spectroscopy feedback lock to the laser frequency. c)~Sketches of the three methods of reducing noise from the seed of the 4WM process: Method 1 - attenuating the amplified seeded beam using a half-wave plate and polarizing beamsplitter to balance the conjugate. Method 2 - the amplified seeded beam balanced by its conjugate and a second seed that goes around the gain. Method 3 - a dual-seeded process, with the two amplified seeded beams and their conjugates balancing on the two detectors.}
\label{fig1}
\end{figure*}

\section{Experiments}
While a 4WM process that generates twin beams that are degenerate in frequency is possible in Rb vapor, the strongest and most easily generated 4WM process is a non-degenerate scheme pumped by a single frequency of light~\cite{ref28}, as indicated in Fig.~\ref{fig1}. A seed at the probe frequency is crossed with the pump beam at a small angle in a vapor cell and generates twin beams (an amplified probe and its conjugate), on opposite sides of the pump. For a seed with a normalized input power of 1 the amplified output probe beam then has a power numerically equal to $G$ (the value of the 4WM gain), and the beam at the conjugate frequency has a power numerically equal to $G-1$. The frequencies of the beams differ by $\approx6$~GHz, and these beams are sent to separate, matched photodiodes in a balanced detector package. The photocurrents are directly subtracted and then amplified, and the noise power is measured with an RF spectrum analyzer. We employ a number of techniques to compensate the DC imbalance of the beams in this configuration and compare the results.

In addition to the intensity balancing techniques, discussed below, the choice of a large phase-matching angle (at the cost of 4WM gain and bandwidth) is important to reduce the collection of scattered pump light. The associated reduction in gain can be recovered to some degree with increased cell temperature (i.e., increased atomic density). With non-integrating detectors the inclusion of a delay line is necessary to compensate for the differing group velocities of the probe and conjugate beams, particularly at higher measurement frequencies~\cite{ref26, ref29}. Finally, the locking and frequency-narrowing of the diode seed laser is also important. Using these techniques together we have observed sub-shot-noise intensity-difference signals at frequencies below 10 Hz, and more than 5 dB of squeezing at 20 Hz and above.

The 4WM process and experimental set-up that we use is sketched in Fig.~\ref{fig1} and is similar to that of Refs.~\cite{ref26, ref28}. A grating-tuned diode laser emitting 90~mW of power provides the seeds for the tapered-amplifier (pump) as well as for the 4WM ``probe'' beams. Approximately 30 mW of light from the diode is used to seed a 2 W tapered amplifier. This light is sent through an optical fiber to create a pump beam of $\approx1.5$ mm $1/e^2$ diameter with 750~mW of power that is detuned in the range of $\Delta=800$~MHz to 1.3~GHz to the blue of the S$_{1/2}$ (F=2) $\rightarrow$P$_{1/2}$ (F=3) transition in $^{85}$Rb. The Rb cell is 1.2~cm long and is heated to a temperature of approximately 120~$^\circ$~C. A pair of seed beams at the probe frequency can be introduced on either side of the pump, at a small angle (approximately 0.3 degrees to 0.5 degrees) relative to the pump beam which allows phase-matching of the 4WM process. The probe seed beams have a $1/e^2$ diameter of 0.55~mm and are derived from the diode laser beam by double-passing a fraction ($\approx45$~mW) of its output through a 1.5-GHz acousto-optic modulator (AOM), resulting in a stable 2-photon detuning of $\delta=-2$~MHz for the process. The detectors have a quantum efficiency of $\approx95\%$ at 795~nm.

Most of the experiments using this 4WM system have been performed using Ti:sapphire lasers because of their superior beam quality and noise properties. Diode laser/tapered amplifier systems, however, will most likely be required for this sort of squeezing to become portable and useful outside of the lab. Unfortunately, these systems are especially noisy (above shot noise) at low frequencies. In particular, if one derives the probe-frequency seed beam from the output of a tapered amplifier it can have relatively large amounts of technical noise. In our case, we derive our seed beam(s) by frequency-shifting part of the direct output of our diode laser with a double-passed AOM and coupling into a fiber. Inevitably extra noise, particularly at low frequencies, is also introduced onto the probe seed beam(s) by the RF amplifier and beam-pointing instabilities caused by the AOM and the fiber coupling.

The diode laser can be frequency-narrowed by locking it to a polarization spectroscopy signal in a warm Rb vapor reference cell~\cite{ref30}. This stabilizes the laser against drifts but also narrows the laser linewidth from $\approx$ 200 kHz to <10 kHz, full width at half-maximum (FWHM). This locking significantly affects the low-frequency squeezing performance, as described below.

\begin{figure}
\centering
\includegraphics[scale=0.2]{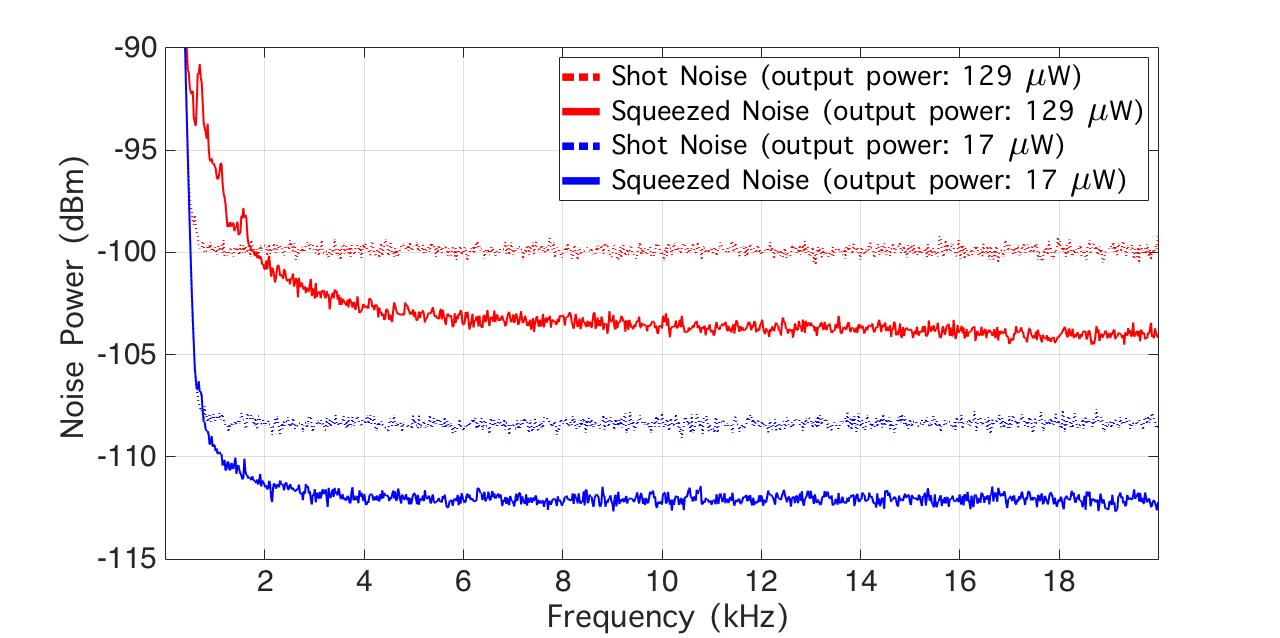}
\caption{Noise power versus measurement frequency for different powers in the seeding beams. The  4WM gain $\approx20$ in each case (pump power = 420~mW, cell temperature=128~$^\circ$~C, $\Delta=1.2$~GHz, $\delta=-2$~MHz, pump-probe angle of 0.5(1) degrees). The upper (red) curves show the measured shot noise and twin-beam intensity-difference squeezing for a seed power of 6.4 $\mu$W (output power $\approx129$~$\mu$W) and lower (blue) curves show the shot noise level and intensity-difference squeezing for a seed power of 0.9~$\mu$W (output power $\approx17$~$\mu$W). The resolution bandwidth (RBW) is 100~Hz and the video bandwidth (VBW) is 1~Hz for these measurements. The output beam powers here are unbalanced due to the injected seed beam.
For equivalent conditions otherwise, the lower seed power results in the squeezing being extended to lower frequencies. The electronic noise is about 20 dB below the shot noise level and is not subtracted from these traces.}
\label{fig2}
\end{figure}

If we input a single seed beam into the 4WM cell we obtain probe and conjugate beams with strong quantum correlations in the MHz frequency range. Our limitation on the squeezing at low frequencies results from the imbalance caused by the presence of the seed. At DC the seeded and unseeded beams will have different intensities, and the spectral powers of the two beams will generally be unbalanced at low frequencies as the gain fluctuates. This imbalance is partially offset by some amount of absorption at the probe frequency (as labelled in Fig.~\ref{fig1}(a)), because it is closer to the atomic resonance, but there is always an imbalance at DC. Attenuating the amplified seeded beam to balance the intensities is effective, however this will also reduce the measured squeezing if the seed beam is not shot-noise-limited. This can be a particular problem if there is a large amount of gain noise at low frequencies. (In severe cases, when the noise becomes larger than the common-mode rejection of the balanced detector, the subtraction will be imperfect.) We show in Fig.~\ref{fig2} how the low-frequency limit of the squeezing is improved for a lower seed power, with otherwise identical conditions.

There are a number of techniques that can be used to balance the power in the two beams. We will show results from the three approaches indicated in Fig.~\ref{fig1}(c). The simplest method (Method 1) is to attenuate the probe (seeded) beam. While this works to balance the average powers, it also slightly attenuates the correlations between the beams, thus reducing the overall squeezing level. In addition, if there are fluctuations in the gain the fixed attenuation cannot compensate for them. A second approach (Method 2) is to generate two beams at the probe frequency from the same source, but only using one to seed the 4WM process. The amplified probe is sent to one photodiode, while the other beam is sent around the gain region in the cell and is brought to the other photodiode directly, along with the conjugate of the amplified probe. The probe-frequency seeds, in the absence of gain, can be balanced at the shot-noise level, while the light generated by the 4WM gain, at both probe and conjugate frequencies, is balanced at a sub-shot-noise level. (Absorption in the vapor complicates this picture.) Finally, a third approach (Method 3) is to generate two seed beams and send equal amounts of probe light through the gain and onto each photodiode. This creates two pairs of twin beams, resulting in each detector seeing the two frequencies of light from one probe and one conjugate beam.  (The probe and conjugate frequencies are sufficiently far apart ($\approx6$~GHz) that the beat note is not observable with the detectors used here.) The stimulated portions of the output beams (with power equal to $G-1$ at each frequency and in each direction) are matched at sub-shot-noise levels on the two detectors, while the injected seed portions of the beams at the probe frequency are balanced on the detectors at the level of shot noise for the seed power.  In Sec. 3, we analyze how these methods compare for generating squeezing at low frequencies.

\section{Results}
Two observations are important for the achievement of squeezing at very low frequencies. The first is that, while the angle between the seed and the pump partly determines the gain through the phase matching conditions, it also partly determines how much scattered pump light is collected by the photodetectors. Since the scattered pump light contains a lot of low frequency noise, we observe an improvement in the low-frequency squeezing by increasing the angle between the pump and probe beams beyond that for optimal gain. The pump light is polarized orthogonal to the probe and conjugate beams at the input and is largely deflected by the polarizing beam splitter at the output of the 4WM cell. A small amount of pump light, however, is depolarized and scattered, primarily at small angles, and is collected by the detectors. This contributes a large background at low frequencies, when compared to the signal level. 

\begin{figure}
\centering
\includegraphics[scale=0.2]{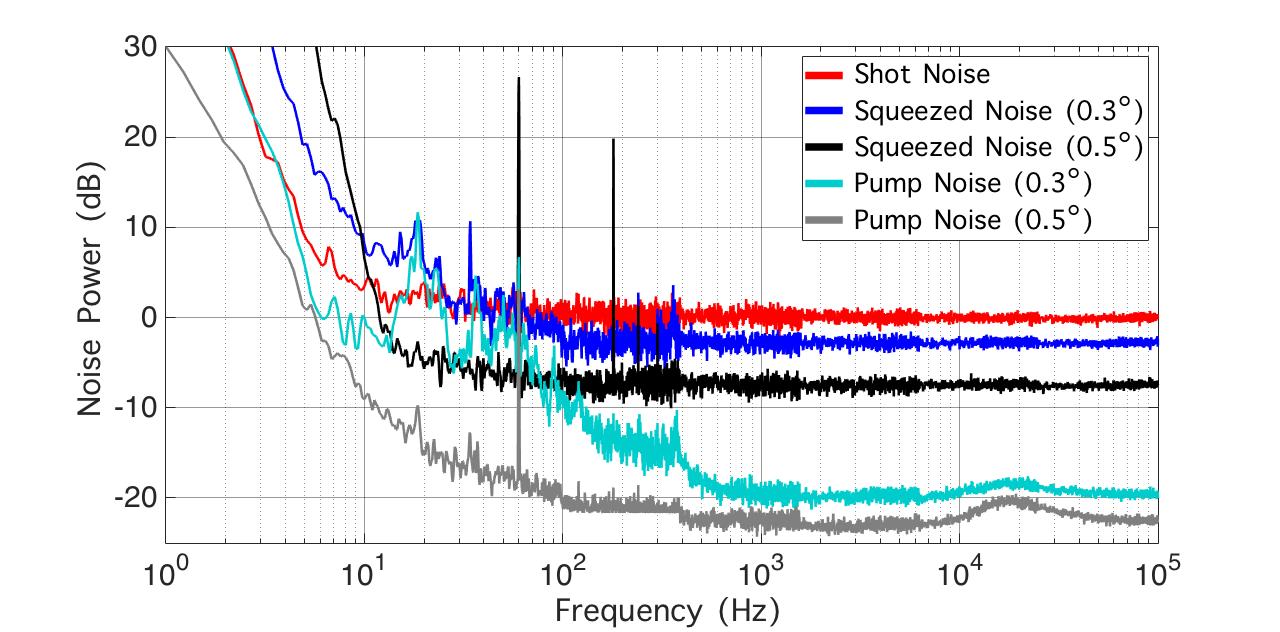}
\caption{Spectra of shot noise, squeezing, and scattered pump light versus frequency, comparing two different phase-matching angles. In this case the 4WM process is dual-seeded (Method 3). The seed laser is locked at a one-photon detuning of 1.3 GHz and frequency-narrowed in each case, and no delay lines are inserted in any of the beams for these measurements.  The two-photon detuning is kept at -2 MHz and the pump power is 750 mW. The plotted spectra (from the top at high frequency) are the a) shot noise measurement (red); b) intensity-difference squeezing with a phase-matching angle of 0.3(1) degrees, gain of 5, and cell stem temperature of 91~$^\circ$~C (blue); c) intensity-difference squeezing with a phase-matching angle of 0.5(1) degrees, gain of 10, and cell stem temperature of 97~$^\circ$~C (black); d) collected scattered pump light for an angle of 0.3(1) degrees (cyan); e) collected scattered pump light for an angle of 0.5(1) degrees (grey). All uncertainties are one standard deviation statistical uncertainties unless otherwise noted.  The electronic noise is not subtracted from these traces, which limits the lower pump noise trace above 100 Hz. The RBW is 0.24 Hz for frequencies below 400 Hz, and progressively larger for higher frequencies, up to 62 Hz above 6 kHz. }
\label{fig3}
\end{figure}

At the cost of smaller gain, this scattering can be reduced by increasing the angle between the probe seed and the pump beam. In Fig.~\ref{fig3}  we show the power spectrum of  the scattered pump light along with the squeezing spectrum for two pump-probe angles. The typical optimal phase-matching angle is $\approx0.3$~degrees Ref.~\cite{ref26}. By increasing this angle to $\approx0.5$~degrees (at a cost of reducing the gain and the gain bandwidth) the scattering background can be reduced by enough to make a difference in the measured squeezing at very low frequencies.  (The gain can be recovered by increasing the temperature of the Rb cell or changing the pump detuning, although absorption losses in the Rb vapor are increased as well.) It is apparent in Fig.~\ref{fig3} that the scattered pump light limits the low-frequency squeezing for the smaller phase-matching angle and that the low-frequency squeezing can be improved by increasing the pump-probe angle.  We note that scattered pump light is not a concern for experiments employing homodyne detection, where interference with the local oscillator effectively provides a narrowband frequency filter, but it does affect direct intensity detection measurements. 

\begin{figure}[!t]
\centering
\includegraphics[scale=0.4]{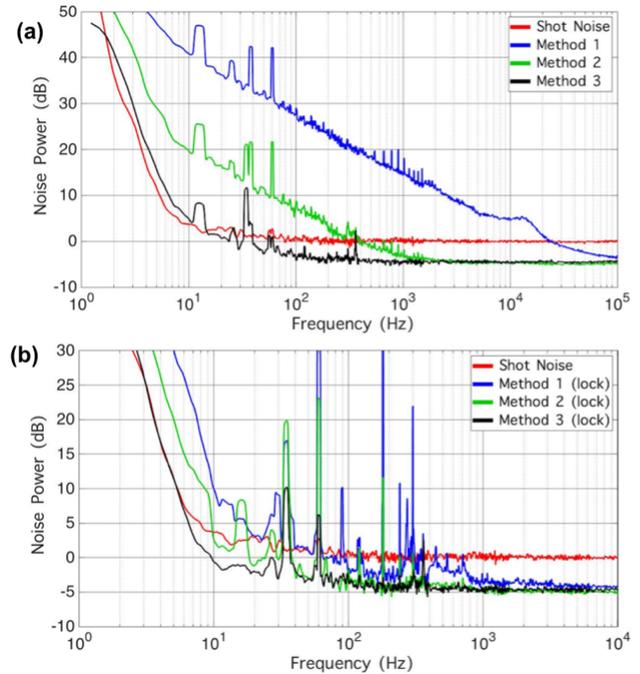}
\caption{Spectra of shot noise and intensity-difference squeezing versus frequency, comparing seeding techniques and laser locking. The 1-photon detuning is 1.3 GHz, the gain is 10, the cell stem temperature is 97~$^\circ$C, and the phase-matching angle is 0.5(1) degrees. In (a) the seed laser is unlocked, and in (b) the laser is locked and frequency-narrowed. In each case the traces are the shot noise measurement (red); intensity-difference squeezing with an attenuated probe beam (Method 1, blue); intensity-difference squeezing with an extra probe seed added to the conjugate beam at the detector (Method 2, green); intensity-difference squeezing with dual probe seeds (Method 3, black). There is a delay line for the conjugate beams in each case, which was included for ease of combining beams on the detectors. The measurements were taken with a box enclosing the beam paths after the 4WM cell. The electronic noise, about 20 dB below the shot noise, is not subtracted from these traces. The RBW is 0.24 Hz for frequencies below 400 Hz, and progressively larger for higher frequencies, up to 62 Hz above 6~kHz. A running average over 11 points is used to smooth the data.  Note the change of scale on the axes between panels (a) and (b).}
\label{fig4}
\end{figure}

The second important practical observation is that, as discussed above, for a single probe-conjugate pair it is the (unbalanced) injected seed light that is one of the major limitations to the squeezing. We see all of the low-frequency technical noise on the seed (if it is not at the shot noise limit) reproduced in the squeezing spectrum, as well as noise due to the changing imbalance caused by gain fluctuations in the 4WM. The squeezing can generally be improved by intentionally attenuating the stronger, seeded (probe) beam~\cite{ref25} (Method 1). Gain fluctuations will still limit the performance, as this balancing can be optimized for only one fixed value of G, and the gain fluctuations will always result in some imbalance, and consequently excess noise. Since there tends to be more noise in the gain at low frequencies than at higher frequencies, this tends to limit the squeezing in that regime first. 
In Fig.~\ref{fig4} we compare the three approaches to balancing the beams for the two situations of the seed laser being unlocked (Fig.~\ref{fig4}(a)), and the laser being locked and frequency-narrowed (Fig.~\ref{fig4}(b)). 

Method 2, described in Sec. 2, allows one to overcome certain sources of noise in the seed beam by balancing part of the power of the amplified probe beam on one photodiode with that of a separate, un-amplified seed beam derived from the same source (along with the conjugate beam) on the other photodiode.  The results of this approach are also shown in Fig.~\ref{fig4}, and it is clear that this technique provides a substantial improvement over Method 1, particularly for the unlocked laser. This method allows for nearly balanced powers on each photodiode along with a straightforward means to cancel any classical noise on the seed beam itself. However, this method does not provide a means for canceling any noise that arises due to fluctuations in absorption of the probe seed in the gain region of the vapor cell.

Method 3 provides a means to overcome excess noise due to fluctuations in the 4WM gain and the absorption. As shown in Fig.~\ref{fig1}(c) Method 3 involves two 4WM processes (1 and 2) generated by the dual-seeding. We detect probe 1 and conjugate 2 on one photodiode, and probe 2 and conjugate 1 onto the second photodiode, so that each diode sees a power of $2G-1$ times the input seed power(s). The seeds now balance to the shot noise level, even with noisy bright seeds and relatively low gain, and the amplified portions of the probe beams can balance to well below the shot noise level with the generated conjugate beams. The result of introducing the two probe-frequency seed beams simultaneously in this way is shown in Fig.~\ref{fig4}. The low-frequency squeezing is noticeably improved beyond Methods 1 and 2 by balancing the beams in this way. 

In order to achieve the best low-frequency results the seed laser itself needs to be stable. The polarization spectroscopy lock of the diode laser indicated in Fig.~\ref{fig1}(b) stabilizes the drift of the diode laser frequency over the long term but, given that its detuning from the atomic resonance is $\approx$ 1300 MHz, the stabilization of this 1-photon detuning is probably not crucial to the squeezing performance. The 2-photon detuning and the coherence of the seed are established by an RF source and an AOM, and are also quite stable. In our experiments, however, the lock also takes out many fast frequency fluctuations and narrows the laser to < 10 kHz linewidth, and we feel that this is the important feature of the lock. In Fig.~\ref{fig4} we show how this laser lock impacts the intensity-difference squeezing at low frequencies.  It is clear that locking is particularly important at the lowest frequencies.

A complication of the non-degenerate 4WM scheme is that the group velocities in the vapor cell at the probe and conjugate frequencies are different, leading to intensity fluctuations in the conjugate beam running ahead of those in the probe~\cite{ref29}. This results in the need for a delay line in the conjugate beam of about 10 ns, depending on the conditions (in particular the 2-photon detuning), in order to optimize the squeezing~\cite{ref26}. This primarily affects the squeezing observed at higher measurement frequencies.  Delay lines are only included in the present experiments to facilitate combining the multiple beams on the photodetectors.

Finally, beam jitter, or beam-pointing instabilities can couple noise into the present measurements in several different ways. In addition to the beam pointing coupling through the varying efficiency of the AOM, or via optical fiber coupling, the non-uniformity of the photodiodes themselves can have an important effect. McKenzie et al.~\cite{ref31} observed that below about 200 Hz the conversion of beam motion on the diode into apparent intensity noise becomes significant. A box was placed around the beams after the 4WM cell in our apparatus to protect the beams from air currents in the laboratory and was required for the consistent observation of squeezing at frequencies below about 20 Hz.

\section{Discussion}
Careful power balancing, laser locking/narrowing, and a large phase-matching angle all contribute to being able to observe squeezing down to low frequencies. We have obtained squeezing down to frequencies < 10 Hz, limited only by the general stability of the optics and laboratory environment. The 4WM process is not fundamentally limited at low frequencies, and the single-pass-gain optical system used here does not couple acoustic noise into the measurements as a build-up cavity would. The pointing instability of the beams due to air currents surrounding the warm vapor cell seems to be a major contributor to the remaining low-frequency limitations in the system. 

The input seed beams for the 4WM processes are constructed with a simple beamsplitter, and thus they have the identical spatial mode structure. The output beams will also reflect the same spatial mode structure, although the beams will perhaps be somewhat distorted by Kerr lensing in the gain medium. The distortions can, in principle, be corrected for, and the intensity subtractions can be made mode-by-mode. Perhaps a more straightforward method would be to employ flat-top beam-shaping optics for the pump, which could eliminate the lensing and reduce the Kerr effect here to uniform phase shifts across the beams. In that case, with a careful alignment, the probe and conjugate beams could be overlapped by simply seeding the second probe beam along the identical path as the conjugate beam from the first seed. In either case, sub-shot-noise absorption imaging should be possible.

Such imaging would be straightforward for broadband absorptive or opaque objects, and even for refractive objects that are far enough off-resonant that the frequency difference (6 GHz in our case) between the probe and conjugate beams is unimportant. On the other hand, one feature of the 4WM generation scheme that is attractive for scientific applications is that it produces narrowband light that can, for example, be used for coupling to cold atoms. In the present configuration the beams are composed of two pairs of non-degenerate, narrowband, multi-spatial-mode twin-beams. In spite of the dual-frequency nature of the beams, they can be adapted to imaging applications, even for narrowband absorbers. One could, as described above, overlap these beams, but then only approximately half of the light would interact with the atoms. Alternatively, one can separately detect the two probe beams (P1 and P2) and conjugate beams (C1 and C2) on a CCD detector, align the beams in software to match the spatial modes, and group them as follows: $\text{C1}=\text{P1}+\text{C2}-\text{P2}$. This results in a single-color imaging beam (C1) that can be referenced to the remaining combination of beams that eliminates the noise on the seed beams. While any loss in the optical system will limit the ability to observe squeezing, CCD cameras with quantum efficiencies of >95\% at the Rb D1 line of 795 nm are available commercially. The consequences of the dual-seed scheme for experiments performed with homodyne detection should be relatively limited. In the case of dual-seeding a bi-chromatic local oscillator would be required~\cite{ref32}, and the relevant noise signals should simply add.

Finally, if we pump the 4WM process near the frequencies of the probe and conjugate beams indicated above, we can generate single-mode squeezing~\cite{ref33}, or we could generate twin beams at the same frequency that are not degenerate in direction~\cite{ref34}. In that case we would not have to use bi-chromatic local oscillators for homodyne detection, but we would have a phase-sensitive process where the twin seeds would have to be carefully phase-stabilized with respect to each other. The present two-mode squeezing results show that the 4WM process is not intrinsically limited at low frequencies, and imply that this sort of phase-sensitive configuration should be capable of generating single-mode squeezing at similarly low frequencies, regardless of whether or not the process is seeded.

\section{Conclusion}
Using a bright-beam 4WM scheme in warm Rb vapor we have been able to achieve intensity-difference squeezing of more than 5 dB down to frequencies below 20 Hz. Experiments that have made the most progress thus far toward quantum-correlated imaging have typically used unseeded nonlinear processes.  Our work presents a means to overcome technical limitations associated with the use of a seed for bright-beam quantum imaging. The lack of a resonator, which is susceptible to acoustic noise coupling, makes these single-pass 4WM schemes capable of generating strong squeezing at very low frequencies, and we have demonstrated this by balancing the seed noise in the detection. Our optimized 4WM scheme, with intensity balancing, an increased phase-matching angle, and the locking/frequency-narrowing of the seed laser, results in the ability to generate intensity-difference squeezing down to frequencies where we are limited by detector electronic noise or by simple beam-pointing instability. This should allow the application of squeezed light to a number of practical applications that demand low frequency measurements. It should also be useful for coupling to cold atoms or to systems such as optical memories based on electromagnetically-induced transparency~\cite{ref14}.

\section*{Acknowledgements}
Bonnie L. Schmittberger (bschmittberger@mitre.org) - The author's affiliation with The MITRE Corporation is provided for identification purposes only, and is not intended to convey or imply MITRE's concurrence with, or support for, the positions, opinions or viewpoints expressed by the author.

\section*{Funding Information}
Air Force Office of Scientific Research (FA9550-16-1-0423).

\bibliography{LowFrequencySqueezing}

\ifthenelse{\boolean{shortarticle}}{%
\clearpage
\bibliographyfullrefs{sample}
}{}


\end{document}